\documentclass[aps,prb,twocolumn,superscriptaddress,groupedaddress,floatfix]{revtex4-1}
\usepackage{amsmath}
\usepackage{amsfonts}
\usepackage{amssymb}
\usepackage{graphicx}
\usepackage{bm}
\usepackage{color}
\usepackage[normalem]{ulem}

\begin{document}

\title{Excitonic mass gap in uniaxially strained graphene}
\author{Anand Sharma} 
\email{sharma@itp.uni-frankfurt.de}
\affiliation{Department of Physics, University of Vermont, 82 University Place, Burlington, Vermont 05405, USA}
\affiliation{Institut f\"ur Theoretische Physik, Universit\"at Frankfurt, Max-von-Laue Str. 1, 60438 Frankfurt, Germany}

\author{Valeri N. Kotov}
\affiliation{Department of Physics, University of Vermont, 82 University Place, Burlington, Vermont 05405, USA}

\author{Antonio H. Castro Neto}
\affiliation{Centre for Advanced 2D Materials and Graphene Research Centre, National University of Singapore, 6 Science Drive 2, Singapore 117546}

\begin{abstract}
We study the conditions for spontaneously generating an excitonic mass gap due to Coulomb interactions between anisotropic Dirac fermions in uniaxially strained graphene. The mass gap equation is realized as a self-consistent solution for the self-energy within the  Hartree-Fock mean-field and static random phase approximations. It depends not only on  momentum, due to  the long-range nature of the interaction, but also on the velocity anisotropy caused by the presence of uniaxial strain. We solve the nonlinear integral equation self-consistently by performing large scale numerical calculations on variable grid sizes. We evaluate the mass gap at the charge neutrality (Dirac) point as a function of the dimensionless coupling constant and anisotropy parameter. We also obtain the phase diagram of the critical coupling, at which the gap becomes finite, against velocity anisotropy. Our numerical study indicates that with an increase in uniaxial strain in graphene the strength of critical coupling decreases, which suggests anisotropy supports formation of excitonic mass gap in graphene.
\end{abstract}
\date{\today}


\maketitle

\section{\label{sec:int} Introduction}\indent During the past decade the remarkable physical phenomena demonstrated by graphene\cite{castroneto}, a two-dimensional sheet of carbon atoms arranged in a honeycomb pattern, not only inspired a great deal of fundamental research in other novel two-dimensional materials\cite{bhimanapati} but also lead to significant advances in many promising graphene-based modern technological applications like transistors\cite{schwierz}, optoelectronics\cite{bonaccorso}, sensors\cite{hill,liu}, membranes\cite{nair}, nanocomposites\cite{chee}, supercapacitors\cite{elkady}, and many more\cite{editorial}. However, due to lack of finite spectral gap at the charge neutrality (Dirac) point, this zero-overlap semimetallic material cannot be directly used for nanoelectronics applications\cite{hertel}. \\
\indent The striking physical properties, in particular the electronic properties due to gapless linear low-energy dispersion relation, of this extraordinary material are governed by the chiral (related to the sublattice and time-reversal) symmetry which emerges as a result of the two-dimensional bipartite honeycomb lattice. This chiral symmetry, or {\it{handedness}}, is not only ubiquitous to graphene but also observed in other condensed matter systems\cite{qi}, high-energy physics\cite{cheng1}, chemistry\cite{cotton} and biology\cite{guijarro}. The spontaneous breakdown of this chiral symmetry in graphene corresponds to the generation of spectral or mass gap with the realization of massive Dirac fermions at the Dirac point and paving the way to potential large-scale applications in nano-science. \\
\indent There have been a considerable number of exciting proposals to generate the mass gap or to observe spontaneous chiral symmetry breaking (CSB), which is also termed as semimetal to insulator or excitonic transition, in graphene. One can realize such broken chiral symmetry phases in the presence of an external magnetic field\cite{semenoff} which is similar to magnetic catalysis in quantum electrodynamics ({\it{QED}}$_{2+1}$)\cite{gusynin}, graphene on dielectrics which breaks the sublattice symmetry\cite{zhou,song,nevius}, confining the motion of the charge carriers in graphene quantum dots\cite{ponomarenko} or nanoribbons\cite{han}, interaction-induced localization of charge carriers in the presence of adatoms\cite{rappoport} or vacancies\cite{ulybyshev1}, applying structural changes (axial strain)\cite{pereira1,guinea,downs} or due to electron-electron interactions\cite{khveshchenko,son,herbut,drut,gamayun,armour,sabio,wang1,wang2,wang3,ulybyshev2,popovici,smith,katanin}. While each of these approaches have their own merits, it is very intriguing to understand the consequence of their interplay on the broken symmetry phases. \\
\indent Theoretical studies\cite{park,pereira2,choi} have shown that strain can significantly alter the electronic band structure of graphene i.e., change its noninteracting single particle properties and create anisotropic Dirac fermions which are also found in other solid state systems\cite{tajima,zhang1,virot,moon,feng,yang,zhang2}. Due to such a directional dependent nature of charge carriers, the anisotropic Dirac fermions can find possible applications in low-dimensional devices utilizing anisotropic charge transport and therefore it is of utmost importance to understand their basic properties. Experimentally the strain-induced anisotropic Dirac fermions in graphene have also been examined\cite{ni,mohiuddin,cheng2,zhang3}. But a systematic study of applying, controlling and measuring an axial strain in a monolayer freely suspended graphene have met with serious challenges due to the lack of proper suspension of large micrometer-sized graphene flakes over a trench or due to absence of an efficient method in clamping the graphene samples. Despite these difficulties, very recently, there has been a great deal of progress in achieving tensile strain of up to 14$\%$ using MEMS technology~\cite{garza} or marginal yet well controlled uniform uniaxial strain\cite{polyzos} of nearly 0.8$\%$. 
It is well understood that strain in graphene is of considerable importance\cite{si,amorim} to systematically study the mechanical strength of this atomic thick single layer of carbon atoms in order to be able to use it in developing stretchable, transparent, and carbon based nanoelectronics devices.\\
\indent In the present work, we shall consider the combined effect of uniaxial strain (the simplest possible axial strain that can be applied) and long-range instantaneous Coulomb interaction (without any retardation effects) in breaking the chiral symmetry in graphene. More specifically our goal is to simplify the self-consistent mass gap equation as much as possible but use large scale numerical simulations to elucidate, under certain physical approximation, whether uniaxial strain enhances or suppresses the interaction-induced excitonic mass gap in freely suspended graphene\cite{wang4,tang}. \\
\indent The rest of the paper is organized as follows. In Sec.~\ref{sec:form} we present the mass gap equation, which signifies the spontaneous generation of finite mass or opening of a gap at the Dirac point in graphene. Such a gap equation is realized in the presence of a uniaxial strain, which makes the electronic dispersion anisotropic, and within Hartree-Fock mean-field as well as static random phase approximation for the effective interaction. Due to exceptional difficulty of finding an exact analytic solution of the gap equation, in Sec.~\ref{sec:nr} we present results of our large scale numerical calculations for solving the anisotropic gap equation on a finite sized grid. Using standard numerical methods we obtain the value of the gap at the charge neutrality point as a function of dimensionless effective interaction strength and applied strain. We plot the phase diagram of the critical strength, which is responsible for opening the mass gap, as a function of anisotropy. In the concluding section~\ref{sec:con} we summarize our work.\\

\section{\label{sec:form} Formalism}
\subsection{\label{sec:mod} Theoretical Model} 

\begin{figure}
 \centering
 \includegraphics[height=5.5cm,width=8.25cm]{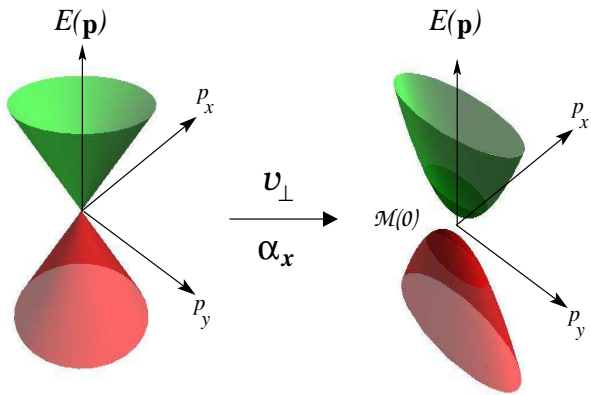}
 \caption{\label{fig:0} Low-energy linear dispersion at the Dirac point in graphene (left) acquires a finite excitonic mass gap, $\mathcal{M}(0)$, as an interplay between externally applied uniaxial strain, $v_{\perp}$, and electron-electron interaction, $\alpha_{x}$.}
 \end{figure}

\indent With an aim to understand the interplay between anisotropic noninteracting dispersion, due to an externally applied uniaxial strain, and electron-electron interaction in graphene we consider a model consisting of Dirac fermions interacting via long-range Coulomb interaction on a uniaxially strained two-dimensional bipartite honeycomb lattice. The essential details of the microscopic model are provided in one of our earlier works, see Ref.~[\onlinecite{sharma1}], but in the following we shall briefly mention a few relevant highlights for the sake of brevity and completeness. \\
\indent We suppose a low-energy effective noninteracting Hamiltonian, 
\begin{equation}{\label{eq:1}}
\hat{H}_{0} = v_{x}p_{x}\hat{\sigma}_{x} + v_{y}p_{y}\hat{\sigma}_{y}, 
\end{equation}
where $v_{x}$ and $v_{y}$ are velocities along spatial $x-$ and $y-$ directions respectively. Here, the two-dimensional quasimomentum or wave vector is given by ${\bf{p}} = (p_{x},p_{y})$ while $\hat{\sigma}_{x}, \hat{\sigma}_{y}$ are the well-known (2 $\times$ 2) Pauli matrices along the $x-$ and $y-$ components respectively of the three dimensional Pauli vector, ${\hat{\bm{\sigma}}} = (\hat{\sigma}_{x}, \hat{\sigma}_{y}, \hat{\sigma}_{z})$. The hat over the Latin and Greek symbols signifies two-dimensional matrices written in the basis of sublattices, $A$ and $B$, of the two-dimensional honeycomb structure. Following Ref. [\onlinecite{sharma1}], we shall consider tensile strain applied along the $y-$ direction and define an anisotropy parameter, $\delta$, proportional to the uniaxial strain\cite{pereira1} which is given in terms of anisotropic velocities such that 
\begin{equation}{\label{eq:2}}
\frac{v_{y}}{v_{x}} = v_{\perp} .
\end{equation}
In the unstrained or isotropic limit, $\delta = 0$, the Fermi velocity is given by $v_{F} = v_{x} = v_{y} \simeq 10^{6} m s^{-1}$ while in the extreme anisotropic limit, $\delta = -1$, the two dimensional graphene is reduced to decoupled chains of carbon atoms. We would like to remark that, though the complete anisotropic extremity is a compelling limit\cite{jin}, due to numerical constraints we shall not be discussing it in this work and refrain our analysis in the proximity of such an extreme limit. However, in the light of recent experimental advancement in applying uniaxial strain in graphene\cite{zhang3,garza,polyzos}, it would be intriguing to examine an interaction-driven phase transition or a continuous dimensional crossover in the limit of strong applied uniaxial strain in freely suspended large scale graphene. \\
\indent On diagonalizing the Hamiltonian in Eq.~(\ref{eq:1}) we obtain the anistropic noninteracting energy dispersion, 
\begin{equation}{\label{eq:3}}
E({\bf{p}}) = \pm \sqrt{(v_{x}p_{x})^{2} + (v_{y}p_{y})^{2}} .
\end{equation}
In the case of unstrained graphene the low-energy dispersion is described by an isotropic circular cone, while the uniaxial strain makes the cone anisotropic and elliptical in the energy plane. Apart from the noninteracting term, we also consider interaction modeled by a long-range Coulomb potential given by 
\begin{equation}{\label{eq:4}}
V({\bf{p}}) = \frac{2\pi e^{2}}{\kappa |{\bf{p}}|} ,
\end{equation}
where $\kappa = 1$ is the dielectric constant for the case of free-standing graphene and $\kappa > 1$ for graphene on a dielectric substrate. The strength of interaction is varied by defining the dimensionless parameter, 
\begin{equation}{\label{eq:5}}
\alpha = \frac{e^{2}}{\hbar \kappa v_{F}} ,
\end{equation}
which is the ratio of the interaction or potential energy to that of the noninteracting or kinetic energy. Its value for freely suspended ($\kappa = 1$) graphene is, $\alpha = 2.2$. Since the uniaxial strain is applied along the $y-$ direction it is obvious that $v_{y}$ will decrease due to increasing bond length while $v_{x}$ shall increase because of decreasing bond length along $x-$ direction. Therefore in the presence of any finite strain on freely suspended graphene it is appropriate, in our model, to consider
\begin{equation}{\label{eq:6}}
\alpha_{x} = \frac{e^{2}}{\hbar v_{x} } , 
\end{equation}
as the strength of interaction. We shall suppose $v_{\perp}$ and $\alpha_{x}$, Eqs.~(\ref{eq:2}) and (\ref{eq:6}), respectively, as model parameters in our calculations and study their interplay in generating an excitonic mass gap in graphene as shown schematically in Fig.~\ref{fig:0}.\\

\subsection{\label{sec:met} Methodology} 

\indent With the minimal description of our theoretical model, we begin our analysis within a self-consistent scheme where the CSB order parameter, related to the generation of finite mass gap, appears in the interacting Green's function which is defined as follows :
\begin{equation}{\label{eq:7}}
\hat{\mathcal{G}}^{-1}({\bf{p}},\omega) = \omega\hat{I} - (v_{x}p_{x}\hat{\sigma}_{x} + v_{y}p_{y}\hat{\sigma}_{y}) - \hat{\Sigma}(\bf{p}),
\end{equation}
where $\omega$ is the external frequency, $I$ is a (2 $\times$ 2) identity matrix and the self-energy is given by 
\begin{equation}{\label{eq:8}}
\hat{\Sigma}({\bf{p}}) =  \mathcal{M}({\bf{p}})\hat{\sigma}_{z},
\end{equation}
with $\hat{\sigma}_{z}$ being the $z-$ component of the three dimensional Pauli vector and $\mathcal{M}({\bf{p}})$ is the mass gap. Since $\hat{\sigma}_{z}$ anti-commutes with the Hamiltonian given in Eq.~(\ref{eq:1}), any non-zero value of $\mathcal{M}({\bf{p}})$ will result in opening a spectral gap in the dispersion relation indicating sublattice symmetry breaking, i.e., spontaneous CSB due to electron-electron interaction. It is well known that any finite mass gap can also be realized in the noninteracting limit due to spin-orbit coupling which breaks the sublattice symmetry, but for graphene its value is negligibly small. Experimentally\cite{mayorov} the upper limit for the band gap at low temperatures, set by charge inhomogeneities, is around 1 meV. The anisotropic gapped energy spectrum at any finite quasimomentum now becomes
\begin{equation}{\label{eq:9}}
E({\bf{p}}) =  \pm \sqrt{(v_{x}p_{x})^{2} + (v_{y}p_{y})^{2} + \mathcal{M}({\bf{p}})^{2}}.
\end{equation}
\indent Now the subsequent task is to obtain a self-consistent equation for the mass gap, $\mathcal{M}({\bf{p}})$. The many-body self-energy in the static (zero external frequency) limit is defined as 
\begin{equation}{\label{eq:10}}
\hat{\Sigma}({\bf{p}}) = i \int \frac{d\omega d^{2}q}{(2\pi)^{3}} V^{\textrm{eff}}({\bf{p}}-{\bf{q}}) \hat{\mathcal{G}} ({\bf{q}},\omega),
\end{equation}
where the effective interaction is given by
\begin{equation}{\label{eq:11}}
V^{\textrm{eff}}({\bf{p}}) =  \frac{V({\bf{p}})}{1 - \Pi({\bf{p}}) V({\bf{p}}) }.
\end{equation}
Here, the bare interaction is as defined in Eq.~(\ref{eq:4}) with $\kappa = 1$ for free-standing graphene. Equation~(\ref{eq:10}) represents the perturbative zeroth order (without any vertex corrections) self-energy in GW theory\cite{hedin} with random phase approximation (RPA) for the screened potential. In the absence of screening due to electron-electron correlation the self-energy becomes the simplified Hartree-Fock exchange. This theory can be extended in a straight forward manner by taking into account the wavefunction renormalization along with vertex corrections and by considering the dynamic particle-hole polarization bubble, $\Pi ({\bf{p}},\omega)$, defined in terms of interacting Green's function, Eq.~(\ref{eq:7}), thereby formulating the fermionic and bosonic self-energies in terms of full self-consistent nonperturbative Dyson-Schwinger equations\cite{dyson}. But it becomes a formidable task to perform large scale numerical calculations for those self-consistent self-energy equations. Therefore, in order to simplify the self-consistent mass gap equation and to be able to implement very large grid sizes, we restrict the effective interaction, Eq.~(\ref{eq:11}), to static RPA where the particle-hole bubble is given as
\begin{equation}{\label{eq:12}}
\Pi({\bf{p}}) = -\frac{N_{f}}{16 v_{x}v_{y}} \sqrt{ (v_{x}p_{x})^{2} + (v_{y}p_{y})^{2} },
\end{equation}
with $N_{f}=4$ being the number of fermionic flavors in graphene corresponding to the two sublattice and two valley degrees of freedom. After inserting Eq.~(\ref{eq:7}) in Eq.~(\ref{eq:10}) and performing frequency integration, the self-consistent anisotropic mass gap equation becomes
\begin{equation}{\label{eq:13}}
\mathcal{M}({\bf{p}}) = \int \frac{d^{2}q}{(2\pi)^{2}} V^{\textrm{eff}}({\bf{p}}-{\bf{q}}) \frac{\mathcal{M}({\bf{q}})}{2\vert E({\bf{q}})\vert} .
\end{equation}
On using Eqs.~(\ref{eq:4}), (\ref{eq:9}), (\ref{eq:11}), and (\ref{eq:12}) we can write Eq.~(\ref{eq:13}) in its final form as
\begin{align}{\label{eq:14}}
\tilde{\mathcal{M}}({\bf{p}}) =& \pi \alpha_{x} \int \frac{d^{2}q}{(2\pi)^{2}} \frac{\tilde{\mathcal{M}}({\bf{q}})}{\sqrt{q_{x}^{2} +v_{\perp}^{2}q_{y}^{2} + \tilde{\mathcal{M}}({\bf{q}})^{2} } }  \nonumber \\
& \frac{1}{ |{\bf{p}}-{\bf{q}}| + \frac{\pi \alpha_{x} N_{f}}{8 v_{\perp}} \sqrt{ (p_{x}-q_{x})^{2} +v_{\perp}^{2}(p_{y}-q_{y})^{2}}} ,
\end{align}
where $\tilde{\mathcal{M}}({\bf{p}}) = \frac{\mathcal{M}({\bf{p}})}{E_{c,x}}$ is the scaled mass gap with $E_{c,x} = v_{x} \Lambda_{x} $ being the cutoff energy scale. As seen in Eq.~(\ref{eq:14}) the scaled mass gap, $\tilde{\mathcal{M}}({\bf{p}})$, not only depends on the quasimomentum because of the long-range nature of interaction but also on the anisotropy due to uniaxial strain. We solve Eq.~(\ref{eq:14}) self-consistently on a two-dimensional finite size grid with varying number of grid points and for different values of dimensionless interaction strength, $\alpha_{x}$, and anisotropic velocity, $v_{\perp}$. Also, apart from presenting results for $N_{f} = 4$ we shall also consider the unscreened case, $N_{f} = 0$, which is equivalent to calculating the effective interaction within Hartree-Fock or mean field approximation neglecting the electron-electron correlation effects. Using standard numerical methods, we shall find the extrapolated value for the mass gap, $\tilde{\mathcal{M}}(0) = \frac{\mathcal{M}(0)}{E_{c,x}}$, whose finite value will indicate spontaneous CSB due to electron-electron interaction in a uniaxially strained freely suspended graphene. As mentioned earlier, we shall restrict our analysis in the proximity of extreme limit of applied strain, $\delta > -1$ or $v_{\perp} > 0$. 

\section{\label{sec:nr} Numerical Results}

\indent In order to have sufficiently large quasimomentum resolution, we use a $\textit{scaled}$ polar grid in a way that there are large number of points distributed near the origin of the co-ordinate system compared to the rest of the grid. There are four sets of total number of grid points, $N=18360, 36360, 72360$ and $108360$, which we consider in our simulations with each converged result requiring from a few minutes to several hours of CPU time on a 1.7GHz AMD Opteron. Since the behavior of the mass gap as a function of quasimomentum is known in the case of unstrained graphene\cite{gamayun}, we consider it as an initial function for the mass gap with specific values on each grid point and calculate the new values after performing the numerical integration for each $N_{f}$ and given $\alpha_{x}$ and $v_{\perp}$. We then compare the newly obtained mass gap with that of initially assumed values. Our criterion of convergence is determined when either the required number of iterations have reached a certain maximum value, which in our calculation is as large as 10000, or the relative error, $\mathcal{E}$, in each iteration is sufficiently small and is given by 
\begin{equation}{\label{eq:15}}
\mathcal{E} = \left| \frac{\tilde{\mathcal{M}}_{i}({\bf{p}}) - \tilde{\mathcal{M}}_{i-1}({\bf{p}}) }{\tilde{\mathcal{M}}_{i-1}({\bf{p}})} \right| < 10^{-4} ,
\end{equation}
where $\tilde{\mathcal{M}}_{i}({\bf{p}})$ represents values of the mass gap on each of the grid points at the $i^{th}$ iteration. It is well known that near the fixed converged point the iterative solutions can sometimes get trapped into a limit cycle where the solutions move around the fixed point very slowly. In order to avoid and eliminate such issues related to the limit cycle we use the damping method or the weighted average scheme for convergence\cite{hartree} which can be understood as follows. At a given $i^{th}$ iteration, if the results are not converged, i.e., the required tolerance limit for the relative error is not reached, the calculations proceed to the next iteration by replacing $\tilde{\mathcal{M}}_{i}({\bf{p}})$ with a linear combination of $(1- w) \tilde{\mathcal{M}}_{i}({\bf{p}})$ and $w \tilde{\mathcal{M}}_{i-1}({\bf{p}})$ where $w$ is the damping or weight factor which lies between zero and 1. \\ 
 \begin{figure}
 \centering
 \includegraphics[height=5.5cm,width=8.25cm]{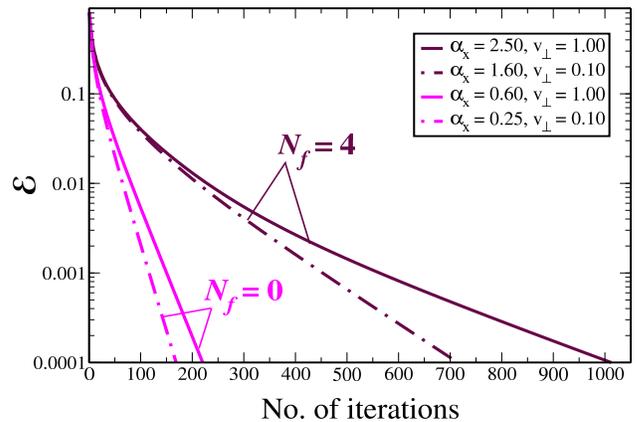}
 \caption{\label{fig:1} Convergence history of the relative error, $\mathcal{E}$, as a function of number of iterations for number of fermionic flavors, $N_{f} = 0 $ and $N_{f} = 4$ for the largest number of grid points.}
 \end{figure}
\indent The computational cost (in terms of iteration number and CPU time) increases rapidly as one approaches the critical value of the coupling constant for each $N_{f}$ and given $v_{\perp}$. In Fig.~\ref{fig:1}, we show convergence history of the relative error, $\mathcal{E}$, as a function of the number of iterations for the number of fermionic flavors, $N_{f} = 0 $ and $N_{f} = 4$ for the largest number of grid points. In the case of $N_{f} = 0$, the computational time for $\alpha_{x} = 0.60$ ($\alpha_{x} = 0.25$) and $v_{\perp} = 1.0 $ ($v_{\perp} = 0.1 $) was 40 (32) CPU hours while, for  $N_{f} = 4$, the computational time for $\alpha_{x} = 2.5$ ($\alpha_{x} = 1.6$) and $v_{\perp} = 1.0 $ ($v_{\perp} = 0.1 $) was 76 (68) CPU hours. Because of the simplified form of the self-consistent equations for the unscreened case $N_{f} = 0$ it is evident that for the given largest number of grid points the number of iterations required to reach the tolerance limit are much lower and the CPU hours are as much as half compared to the screened case $N_{f} = 4$. Moreover, for $N_{f} = 0$ and $N_{f} = 4$, the tolerance limit is reached earlier in the strained case as compared to the unstrained due to the nature of dimensional reduction.\\
\indent For each particular value of $N_{f}, \alpha_{x}$ and $v_{\perp}$, we now consider the mass gap corresponding to the extrapolated smallest quasimomentum, $\tilde{\mathcal{M}}(0) = \frac{\mathcal{M}(0)}{E_{c,x}}$, in each of the given set of grid size. We further extrapolate these values against inverse system size in order to obtain the mass gap value in the continuum limit thereby eliminating finite size effects whose importance has been emphasized in a recent study\cite{smith}. \\
 \begin{figure}
 \centering
 \includegraphics[height=5.5cm,width=8.25cm]{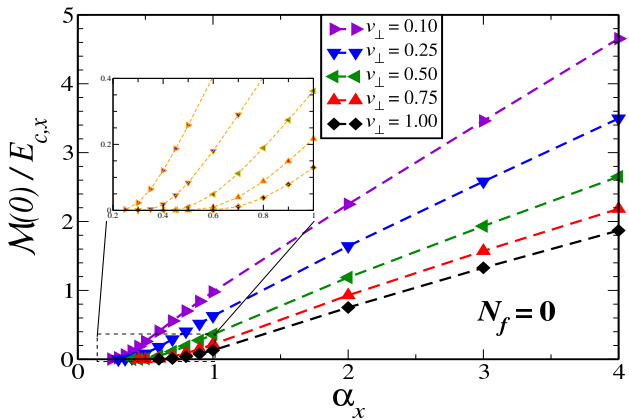}
 \caption{\label{fig:2} Mass gap close to charge neutrality point in strained graphene for number of fermionic flavors, $N_{f}$ = 0, as a function of dimensionless effective interaction strength and for four different values of anisotropy. In the inset, we show the region of the appearance of finite mass gap where the data points are fit to Eq.~(\ref{eq:16}) and shown in broken orange line.}
 \end{figure}
\indent In Fig.~\ref{fig:2} we present the mass gap close to the charge neutrality point in strained graphene for the unscreened case, $N_{f}$ = 0, as a function of dimensionless effective interaction strength and for four different values of anisotropy. For this case, it is apparent from Eq.~(\ref{eq:14}) that for large values of coupling strength the mass gap increases linearly with dimensionless effective coupling constant while for coupling strength near critical values it increases exponentially in accordance with the scaling law\cite{miransky} given by, 
\begin{equation}{\label{eq:16}}
\frac{\mathcal{M}(0)}{E_{c,x}} = A_{0} e^{-\frac{A_{1}}{\sqrt{\alpha_{x} - \alpha_{x}^{c}}}} ,
\end{equation}
where $A_{0}$, $A_{1}$, and $\alpha_{x}^{c}$ are fit to the data points and the fitting curve is shown in the inset of Fig.~\ref{fig:2} as broken orange line. Our result for the critical value within mean-field (unscreened) interaction for the unstrained case, $\alpha^{c} = 0.49$, is comparable to the one reported in Ref.~[\onlinecite{sabio}] using variational method, which in their notation is g$_{c} = 0.5$. \\
 \begin{figure}
 \centering
 \includegraphics[height=5.5cm,width=8.25cm]{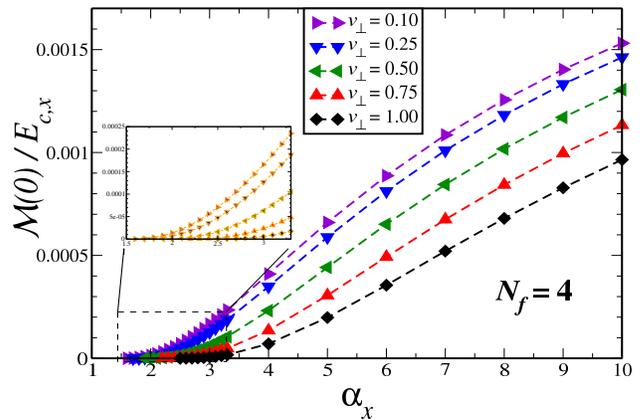}
 \caption{\label{fig:3} Mass gap close to charge neutrality point in strained graphene for number of fermionic flavors, $N_{f}$ = 4, as a function of dimensionless effective interaction strength and for four different values of anisotropy. In the inset, we show the region of the appearance of finite mass gap where the data points are fit to Eq.~(\ref{eq:16}) and shown in broken orange line.}
 \end{figure}
\indent In Fig.~\ref{fig:3}, we again present the mass gap close to the charge neutrality point in strained graphene but for the screened case $N_{f}=4$ within static RPA. We find that for large values of coupling strength the mass gap shows almost sublinear increase or even saturation-like behavior with increasing dimensionless effective coupling constant as opposed to linear increase in the unscreened case. Moreover, it is important to note the values of mass gap for both $N_{f}=0$ and $N_{f}=4$ where the scaled mass gap values are very small in the latter case. These behaviors certainly emphasizes the role of static screening compared to the mean-field approximation. But even for $N_{f}=4$ the coupling strength near critical values shows exponential increase similar to the case for $N_{f}=0$. We again use Eq.~(\ref{eq:16}) to fit the data and the result is shown in the inset of Fig.~\ref{fig:3} as broken orange line.\\
 \begin{figure}
 \centering
  \vspace*{0.5cm}
 \includegraphics[height=5.5cm,width=8.25cm]{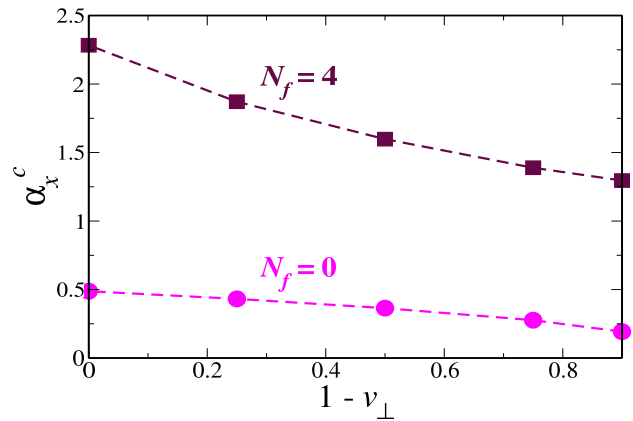}
 \caption{\label{fig:4} Phase diagram showing the dependence of the critical coupling constant on the applied uniaxial strain in graphene.}
 \end{figure}
\indent We now plot the values of the critical coupling constant $\alpha_{x}^{c}$, where the mass gap vanishes for both $N_{f}=0$ and $N_{f}=4$ against the anisotropy parameter as shown in Fig.\ref{fig:4}. We observe that with an increase in anisotropy the value of the critical coupling strength needed to break the chiral symmetry decreases in both the cases with the decrease being larger in the $N_{f}=4$ case. This is in stark contrast to the results obtained by Wang {\it{et al}}~\cite{wang4}, who used six different (static) approximations for the screened interaction including velocity renormalization and obtained non-monotonic dependence of the mass gap on the anisotropy. Moreover, their conclusion is that mass gap always get suppressed as the velocity anisotropy increases; thus the critical coupling increases with increase in anisotropy and therefore anisotropy is not in favor of generating a dynamical mass gap in graphene. On the other hand, our results are in line with the very recent ones reported by Braguta {\it{et al}}\cite{braguta}, where the authors studied similar interplay of fermion velocity anisotropy and long-range Coulomb interaction albeit in three-dimensional Dirac semimetals using Monte Carlo simulations. We also find that within our crude physical approximation for the effective interaction but large scale numerical calculations the value for the critical coupling constant for unstrained graphene, $\alpha^{c} = 2.28$, is larger but very close to its bare value, $\alpha = 2.2$, implying that graphene is in the semimetallic phase. We would like to remark that among other neglected physical aspects in graphene, the velocity renormalization is not only shown to be very important for the correct low-energy description of Dirac fermions\cite{bauer} but is also shown to push the critical coupling to higher values\cite{katanin}. Moreover, it is instructive to comment that in the literature the reported values for the critical coupling constant for the unstrained graphene within various approximations range from 1.1 (Ref.~\onlinecite{drut} using Monte Carlo) to 3.7 (Ref.~\onlinecite{katanin} using functional renormalization group approach). \\

\section{\label{sec:con} Conclusions and outlook}

\indent In summary, we have studied the spontaneous mass gap generation due to long-range Coulomb interactions in  uniaxially strained undoped graphene. We obtain the mass gap equation as a self-consistent solution for the self-energy within Hartree-Fock mean-field and static RPA. The nonlinear integral equation for the mass gap, which depends on the quasimomentum due to the long-range nature of Coulomb interactions and on the anisotropy owing to uniaxial strain, is solved self-consistently by performing large scale numerical simulation on a two-dimensional finite size grid with varying number of grid points. We numerically obtain the mass gap, close to the Dirac point, as a function of the dimensionless coupling constant and anisotropy parameter. The critical coupling, at which the gap becomes finite, is plotted against anisotropy and indicates that with an increase in anisotropy (uniaxial strain) in graphene, the strength of critical coupling decreases which suggests anisotropy supports formation of excitonic mass gap in graphene.\\
\indent Our numerically exhaustive attempt on a simplified version of the model aims towards finding an accurate value of the critical coupling which is responsible for interaction-driven CSB thereby generating an excitonic mass gap in unstrained and strained graphene. Our future approach is to combine nonperturbative methods like the Dyson-Schwinger or functional renormalization group along with large scale numerical calculations. These nonperturbative methods will systematically include dynamic screening\cite{gamayun} with self-consistent addition of mass gap in the polarization bubble\cite{wang1}, velocity renormalization\cite{wang2}, vertex corrections\cite{katanin} and possibly retardation effects\cite{carrington}. 

\section*{\label{sec:ack} ACKNOWLEDGMENTS} 
A.S. and V.N.K. would like to thank A. Del Maestro, C. Herdman and in particular T. Lakoba for fruitful conversations on numerical analysis. 
In addition V.N.K. gratefully acknowledges numerous conversations as well conceptual and technical advice from A. Del Maestro. 
A.S. is grateful to P. Kopietz for his critical reading of the manuscript and for stimulating discussions. This work was supported by the U.S. Department of Energy (DOE) Grant No. DE-FG02-08ER46512. A.H.C.N. acknowledges the NRF-CRP award ``Novel 2D materials with tailored properties: beyond graphene'' (No. R-144-000-295-281). The computational resources provided by the Vermont Advanced Computing Core, which is supported by NASA (Grant No. NNX-08AO96G), are gratefully acknowledged. \\

\end{document}